\documentclass[english,twocolumn,showpacs]{revtex4}
\usepackage{float}
\usepackage{graphicx}
\usepackage{amssymb}
\usepackage{dcolumn}
\usepackage{amsmath}
\makeatother
\usepackage{babel}
\begin{document}

\title{Phase separation in binary colloids with charge asymmetry}

\author{Koki Yoshizawa$^1$,   Nao Wakabayashi$^1$,  Masakatsu Yonese$^1$, Junpei Yamanaka$^1$ and C. Patrick Royall$^2$}
\affiliation{$^1$Faculty of Pharmaceutical Sciences, Nagoya City University, 3-1 Tanabe, Mizuho, Nagoya, Aichi 467-8603, Japan.}
\affiliation{$^2$School of Chemistry, University of Bristol, Bristol, BS8 1TS, UK.}


\email{paddy.royall@bristol.ac.uk}

\date{\today}

\begin{abstract}
We report that binary dispersions of like-charged colloidal particles with large charge asymmetry but similar size exhibit phase separation into crystal and fluid phases under very low salt conditions. This is unexpected because the effective colloid-colloid pair interactions are accurately described by a Yukawa model which is stable to demixing. We show that colloid-ion interactions provide an energetic driving force for phase separation, which is initiated by crystallization of one species.
\end{abstract}


\maketitle

\section{Introduction}

We encounter suspensions of charged colloids every day: their applications range from cosmetics to coatings. Since they are well described by a soft repulsive interaction, charged colloids also form an important model system, and behave in a similar way to materials as diverse as dusty plasmas \cite{book, wysocki2010} and metals \cite{wette2009}. Furthermore, colloidal particles in aqueous suspension form a basic model of biological systems such as proteins \cite{lowen2003}. In all these examples, many-body effects (deviations from density independent pairwise interactions) can be important: many-body effects underly the phase separation we observe here. 

Charged colloids in suspension interact with small ions in the form of added salt, and counterions which balance the charge on the colloids. This system can in principle be described with Coulomb interactions between the small ions and colloids, with an additional hard core term to account for the size of the latter. However, usually the number of interacting species is too large to deal with. The degrees of freedom of the small ions may be integrated out, which leads to an effective colloid Hamiltonian  $W=W_1(\rho_s, \rho_c; {\rho_k})+W_2(\rho_s, \rho_c; {\rho_k}, \{ \mathbf{R} \})+\dots$ where the one-body terms $W_1$ do not depend on the colloid coordinates $\{ \mathbf{R} \}$. Here the counterion number density $\rho_c=\rho_{ion}-2\rho_s$ where $\rho_{ion}$ is the number density of (monovalent) ions, $\rho_s$ is the number density of salt ion pairs and $\rho_k$ is the number density of colloid species $k$ \cite{torres2008pre}. This effective Hamiltonian then depends on state point and is not in general pairwise additive.

Often, one considers the effective colloid-colloid terms $W_2$ only, neglecting higher-order and one-body terms, as the latter are independent of colloid coordinates. The two-body terms involve an effective interaction between the colloids which can be described by the Derjaguin, Landau, Verwey and Overbeek (DLVO) model \cite{verwey1948}. This assumes that the electrostatic potential is sufficiently weak that linear Poisson-Boltzmann theory can be applied. Neglecting van der Waals effects, the effective colloid-colloid interaction takes a hard core Yukawa form 
\begin{equation}
\beta u_{ij}(r)=\begin{cases}
\text{$\infty$} & \text{$r<\sigma_{ij}$}\\
\text{$(1+\Delta_{ij}) \frac{Z_i Z_j}{(1+\kappa \sigma_{ij}/2 )^2}  \frac{\lambda_B}{\sigma_{ij}}   \frac{\exp[-\kappa(r-\sigma_{ij})]}{r/\sigma_{ij}}  $} & \text{$r\ge\sigma_{ij}$}
\end{cases}\label{eqYuk}
\end{equation} 
\noindent where $\beta=1/k_BT$, $\Delta_{ij}=0$ except in the case of non-additivity (see below), $Z_k$ is the charge on colloid species $k$, $\lambda_B$ is the Bjerrum length,
$\sigma_{ij}=(\sigma_i + \sigma_j)/2$ is the mean of the diameters of colloid species $i,j$ and 
$r$ is the centre separation of the two species. The inverse Debye screening length is denoted by $\kappa=\sqrt{4\pi \lambda _B \rho_{ion}}$. Linear Poisson-Boltzmann theory is only valid for small electrostatic potentials (colloid charges). However, at higher potentials, the pair interaction retains a Yukawa form but with a smaller renormalized (effective) charge \cite{alexander1984}. 
Using the effective charge, a phase diagram can be calculated using Yukawa theory, which accurately reproduces experiment, for parameters comparable to ours \cite{monovoukis1989,toyotama2010}.

Treating charged colloids as a one-component system with Yukawa repulsions (with fixed parameters), one does not expect demixing into colloid-rich and colloid-poor phases, as this would typically require some form of attraction between the colloids.  Experimental observations of colloidal fluid-fluid phase separation \cite{ito1994} therefore attracted much attention. Some explanations invoked higher-order terms where the electrostatic repulsions between two colloids might be screened by an intervening third particle 
 \cite{russ2002}. This `macroion screening' was subsequently experimentally verified \cite{brunner2002}, but is a weak effect. Other possibilities focussed on one-body terms, such as ion condensation, which turn out not to be significant for monovalent ions \cite{linse1999} and `volume' terms, where the entropy of small ions, colloid-ion and colloid self interactions contribute to the free energy \cite{zoetekouw2006}. 

Here we consider a mixture of similar sized colloids with different charges, which raises the additional possibility of non-additivity. Non additivity refers to the case where the cross interaction differs from the mean interaction between each species, for example  in Eq. (\ref{eqYuk}) $\Delta_{ij} \neq 0$ when $i\neq j$ and 0 otherwise. For the nearly size-symmetric, dilute colloidal systems we consider, these non-additivity effects are expected to be small, so Eq. (\ref{eqYuk}) holds with $|\Delta_{ij}| \leq 0.01$ \cite{torres2008JCP,allahyarov2009}, and thus we expect nearly additive Yukawa behavior. Like one-component Yukawa systems, additive binary Yukawa systems are not expected to phase separate \cite{hopkins2006}. Remarkably, our experiments reveal demixing into phases rich in high-charge and low-charge particles respectively. The phase rich in high-charge particles is found to be a colloidal crystal. By including one-body $W_1$ contributions through volume terms \cite{torres2008pre,zoetekouw2006} we identify a driving force for phase separation.

\begin{figure}
\includegraphics[width=80mm]{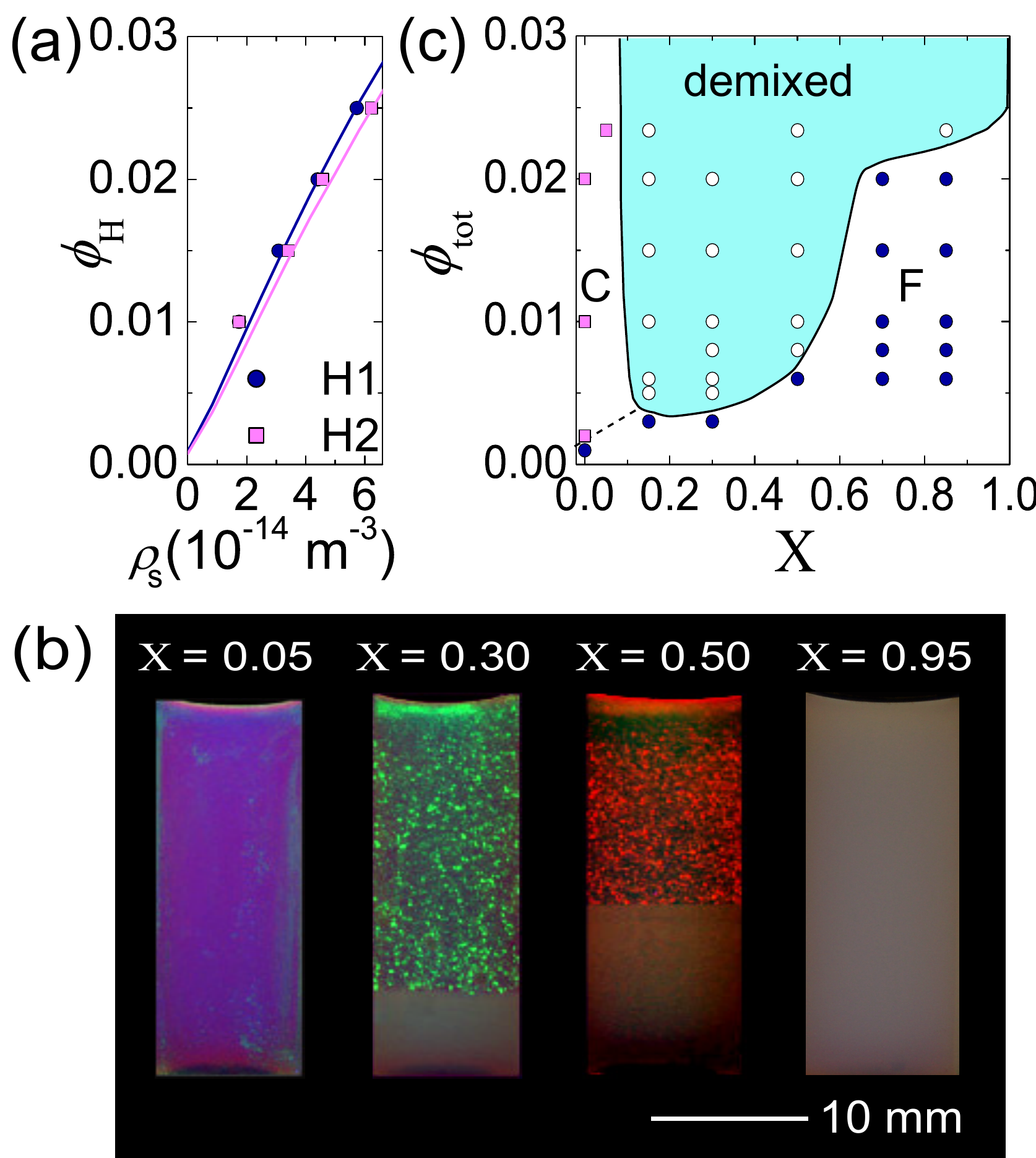}
\caption{(color online) 
(a) Phase boundary in the $(\phi_H,\rho_s)$ plane for the H$_1$ and H$_2$ one-component systems. Experimental results for H$_1$ and H$_2$ are denoted by filled circles and squares respectively. Predictions from Yukawa theory  \cite{robbins1987} are shown as the solid (H$_1$) and dashed (H$_2$) lines. (b) Overiew of phase separation ($\phi_{tot}= 0.0234$, $t = 15$ hrs). From left: $X = 0.05$ (crystal), $0.30$, $0.50$ (phase separated) and $0.95$ (fluid). 
(c) Phase diagram of binary charged colloids defined by $\phi_{tot}$ and $X$ obtained at $t = 240$ hrs. Filled squares, H-rich crystal (C);  open circles, phase separation (demixed); filled circles, fluid (F). Dashed line denotes fluid-crystal phase boundary in the L-dominated regime.}
\label{figPhase}
\end{figure}

\section{Experimental}

We used low charge colloidal silica ($Z = -170, \sigma_L = 110$ nm; species L),  (Japan Catalyst, Co., Ltd). We synthesized higher charge polystyrene particles, H$_1$ ($Z = -870, \sigma_{H1} = 92$ nm) and H$_2$ ($Z = -940, \sigma_{H2} = 90$ nm). Effective charges were determined from electrical conductivity measurements as described in \cite{toyotama2010}. All samples were purified by dialysis and ion exchange \cite{yoshida1999}. Aqueous dispersions of L exhibit no crystallization for volume fractions $\phi_L\leq 0.07$. The phase diagrams for the one-component systems of H$_1$ and H$_2$, in the volume fraction $\phi_H$, salt number density $\rho_s$ plane are shown in Fig. \ref{figPhase}(a) \cite{robbins1987}. For our particles, any deviation from Eq. (\ref{eqYuk}) is expected in the high-charge case. Since our results agree well with the predictions from Yukawa theory, we conclude that it is reasonable here to treat colloid-colloid interactions as a Yukawa system with an effective charge. No significant difference was seen between H$_1$ and H$_2$ which are hereafter termed `H'.

\section{Results}

Phase separation of binary like-charged colloids is shown in Fig. \ref{figPhase}(b). Here the total volume
fraction $\phi_{tot}=\phi_L+\phi_H=0.0234$. We scan from H-rich to L-rich by changing the mixing ratio 
$X=\phi_L/\phi_{tot}$. 
When low charge particles dominate $(X = 0.95)$ (right), the system appears opaque, indicating a colloidal fluid. Meanwhile for a majority of high charge particles $(X=0.05)$ (left), the system is iridescent --- a colloidal crystal. However at $X = 0.3,0.5$ we find a macroscopic phase separation into iridescent (upper) and disordered (lower) regimes with a clear boundary. The change in color indicates that as $X$ increases, so does the lattice constant of the colloidal crystal. The phase diagram at $ t =  240$ hrs is shown in Fig. \ref{figPhase}(c). This has a pronounced asymmetry, such that a higher $\phi_{tot}$ was required for demixing in the case of $L$-rich samples. The phase diagram in Fig. \ref{figPhase}(c) was unchanged up to $t = 1440$ hours. 

\begin{figure}
\includegraphics[width=80mm]{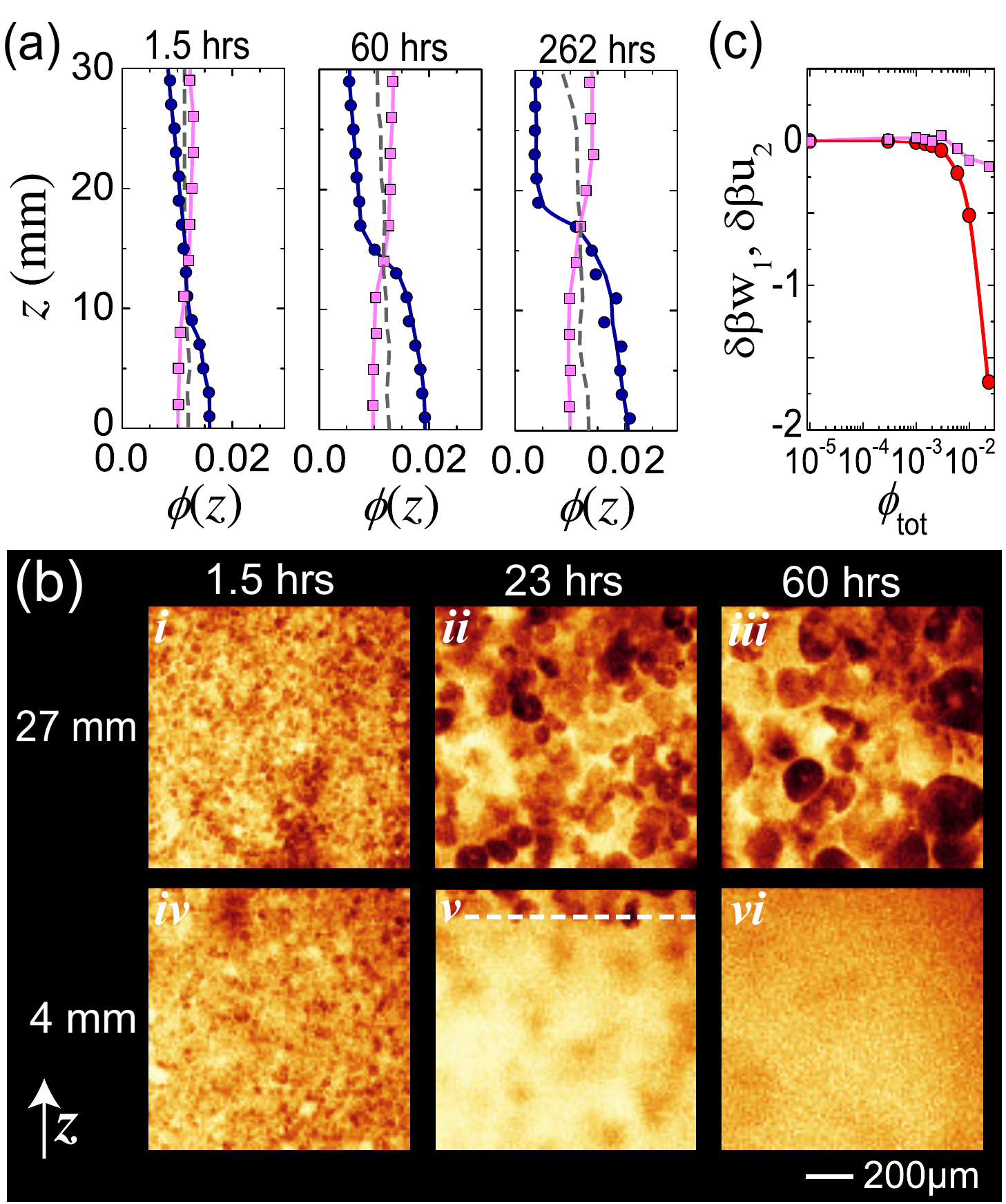}
\caption{(color online)  
(a) Lateral concentration profiles of species L and H$_1$ in their binary mixtures ($\phi_{tot} = 0.0234, X = 0.50$) at $t = 15, 60$ and $262$ hrs. Filled circles, L; open squares, H$_1$. The profiles for L one-component system are also shown for comparison in dashed curves, indicating that sedimentation is not important on these timescales. 
(b) Confocal images of the system immobilized by polymer gel at different times at two heights. Scale bar applies to all images ($\phi_{tot} = 0.0234, X = 0.30$). Dashed line in (b.v) indicates phase boundary. Note that the mixing ratio $X$ is different between (a) and (b). 
(c) Contributions to changes in energy density upon phase separation for a mixing ratio $X=0.5$. Free energy contribution from one-body volume terms $\delta W_1$ (red circles) dominate over two-body potential energy contributions $\delta U_2$ (pink squares).}
\label{figGelled}
\end{figure}

We now turn our attention to the composition of each regime.
Concentration profiles were determined as follows for $L+H$ $(\phi_{tot} = 0.0234, X = 0.50)$. By finding the concentration of silica monomers, (following decomposition with sodium hydroxide) using the molybdenum blue method we obtained $\phi_L(z)$ (silica) at 15 heights $z$. $\phi_H(z)$ (polystyrene) was determined via turbidity measurements in 90\% ethylene glycol whose refractive index is matched to silica. Fig. \ref{figGelled}(a) shows the vertical distribution of L (filled circles) and H (open squares) thus obtained for the samples at $t=1.5$, $60$ and $262$ hrs. The distribution of $L$ becomes inhomogeneous with time due to phase separation, while the H particles show somewhat weaker segregation. At $t$ = 262 hrs, the averaged $\phi_L=0.05$ and $\phi_H=0.014$ in the upper region while in the lower region $\phi_L=0.019$ and $\phi_H=0.010$. 

We then made a  closer investigation of the compositions of the top and bottom regions, using confocal microscopy (Zeiss LSM 510 META). We gelled sucessive samples by curing a UV-sensitive additive  \cite{murai2007} at times $t = 1.5$, $23$ and $60$ hrs. The resulting images at heights $z = 4$ and 27 mm reveal the process of phase separation for ($\phi_{tot} = 0.0234, X = 0.30$) [Fig. \ref{figGelled}(b)]. After $1.5$ hrs, round crystalline domains, a few tens of microns in size, are present throughout the sample. In the top region of the sample, the average domain size then increased to approximately 100 $\mu$m up to $t = 23$ hrs, while the lower region became fluid. After 60 hrs, the top was mostly crystalline domains. If we suppose that the crystals are H-dominated and the fluid which lies between is L-dominated, then the local volume fraction of highly charged particles in the crystals $\phi_H^{loc}<\phi_H(z)$  the mean volume fraction and $\phi_L^{loc}>\phi_L(z)$ where the mean $\phi_{L,H}(z)$ are given in Fig. \ref{figGelled}(a). This indicates that our experiments may be incompletely demixed. The H-rich regions have a microstructure of H crystals and L fluid in the interstices [Fig \ref{figGelled}(b.iii)], while the L-rich fluid also contains H particles.

\section{Discussion}

We now consider a mechanism for the phase separation we observe, recalling that such behaviour is not expected for an additive binary Yukawa system \cite{hopkins2006}. Preferential sedimentation of the denser silica is ruled out: sedimentation of a one-component L system occurs on a timescale of weeks, as the dashed lines in Fig. \ref{figGelled}(a) indicate. By contrast phase separation takes hours. We also tuned the $Z$ value of the L particles by varying the \emph{p}H \cite{yamanaka1998}. For $Z_L = 900$ we observed no phase separation. We conclude that demixing is not related to the different colloid materials, but to the difference in charge between the colloids. 

We identify two mechanisms for phase separation: crystallisation of one species (a two-body effect) and a reduction in potential energy due to ion-colloid coupling (a one-body effect).
Crystal/crystal and fluid/fluid phase separation were not observed. Demixing is then associated with crystallisation of the strongly charged species H, and we believe this is more than coincidental. Crystal nuclei of species H could expel L as the initial phase separation mechanism,  suggesting some asymmetry in the phase diagram as we find [Fig. \ref{figPhase}(c)], along with the absence of phase separation in the case that neither species crystallizes. 
Charged colloids have strong interaction energies and, as mentioned above, additive binary Yukawa systems are stable to demixing \cite{hopkins2006}. 
Moreover entropic contributions from small ions stabilise the mixed state. In other words, without some additional mechanism to lower the free energy, we expect that crystallisation of H would be arrested by the free energy cost associated with phase separation. We therefore enquire as to a mechanism for reduction in free energy, which would promote demixing. 

We proceed by considering the one-body contributions to the Hamiltonian $W_1$. Although these ‘volume’ terms are not related to the colloid coordinates, they capture ionic contributions the free energy. In this way, the  ‘volume’ terms contribute to equilibria between phases with differing compositions. 
In fact, since the ions are so much more numerous than the colloids, they dominate the osmotic pressure, and entropic contributions to the free energy, while ion-colloid coupling is an important part of the potential energy \cite{zoetekouw2006}. 
The contribution to the Helmholtz free energy density from the volume terms reads \cite{torres2008pre}

\begin{equation}
\beta w_1 =\rho_c \left( \ln \frac{\rho_c}{\rho_s}-1 \right) - \frac{1}{2}\sum^2_{k=1}\rho_k \left( \frac{Z^2_k \kappa \lambda_B} {1+\kappa \sigma_k/2} + Z_k \right)
\label{eqW1}
\end{equation}

\noindent where $W_1=Vw_1$ and $V$ is the total volume. The salt number density $\rho_s$ we take from the self-dissociation of water. We compared with higher salt concentrations and found very similar results. We use values close to our experimental parameters with $Z_H = -900$ and $Z_L= -170$, and diameters $\sigma_L =\sigma_H =100$ nm. The osmotic pressure is then $\Pi=\sum^2_{k=1} \rho_k\mu_k-w_k$ where the volume term contribution to the chemical potential is $\mu_k=\partial w_1 / \partial \rho_k$. The Grand Canonical formalism behind Eq. (\ref{eqW1}) includes small ion contributions to $\Pi$ \cite{zoetekouw2006}. Here we assume comlete demixing, so equating the osmotic pressures of the coexisting phases fixes partitioning of volume, which yields the colloid volume fraction of each demixed phase $\phi_H^d$ and $\phi_L^d$. Equation (\ref{eqW1}) then yields $\delta w_1$, the change in free energy density
due to the volume terms upon phase separation. $\delta w_1$ is shown in Fig. \ref{figGelled}(c) for $X=0.5$. For all colloid volume fractions, $\delta w_1$ is negative, suggesting that the system has the potential to demix. However, the magnitude of $\delta w_1$ increases significantly around $\phi_{tot} \sim 0.003$, which is similar to the point at which our experiments demix [Fig. \ref{figPhase}(c)].

Now two-body contributions are also density-dependent. The prefactor and screening length in Eq. (\ref{eqYuk}) both depend on the ionic strength, which is a function of the colloid volume fraction \cite{reneParams}. We therefore also calculate $\delta u_2$, the effective colloid-colloid contribution to the change in potential energy density, using Monte Carlo simulation in the canonical ensemble according to Eq. (\ref{eqYuk}) \cite{MC}. We neglect entropic contributions to the free energy arising from $W_2$, recalling that ionic contributions to $W_1$ are much more significant.
The simulations yield the pressure contributions from 2-body interactions, but these make almost no difference to the values of $\phi_L^d$ and $\phi_H^d$ at coexistence and are neglected. $\delta u_2$ is plotted in Fig. \ref{figGelled}(c). At intermediate densities ($0.001 \lesssim \phi_{tot} \lesssim 0.01$), we find evidence that two-body contributions might suppress phase separation as $\delta u_2>0$. However at higher volume fractions ($\phi_{tot} \gtrsim 0.01$), $\delta u_2$ is dwarfed by the volume term contribution. Our treatment therefore suggests a lowering in free energy upon phase separation as a similar state point to that found in experiments.

Considering the terms in Eq. (\ref{eqW1}), we can identify a driving force for phase separation. The first term represents ion entropy, which tends to oppose phase separation. The second term represents the colloid self-energy, that is to say ion-colloid interactions. Here, partitioning leads to enhanced ion-colloid coupling in the H-rich phase, and we argue that this stablises phase separation. There are other, smaller, effects, for example the change in Debye screening length in each phase following demixing, and its contribution to the potential energy contributions to the colloid-colloid terms $\delta u_2$. We have observed that the lattice constant of crystals of H particles increases with $X$ in Fig. \ref{figPhase}(b). This arises naturally because increasing $X$ at fixed $\phi_{tot}$ reduces the osmotic pressure, allowing the crystals to expand.

Finally, we consider the generality of our findings. The mechanism we propose should be relevant to multicomponent like-charged systems, such as dusty plasmas and protein solutions. Segregation in the latter can have important physiological implications. Using parameters of $\sigma_{ij}=1$ nm and net charges of $Z_H=6$ and $Z_L=1$ and $c_s=0.01$ M to describe proteins in physiological conditions, we find a free energy difference per molecule of $0.1$ $k_BT$ at $\phi_{tot}= 0.13$.

A number of assumptions have been made here, which suggest that our findings could be investigated futher in the future. In particular, (1) we have assumed complete phase separation,  (2) that linear Poisson-Boltzmann theory holds, and (3) that higher order terms such as three-body colloid-colloid interactions can be neglected. For our parameters, for $\phi_{tot}\gtrsim0.01$, (3) appears to hold \cite{russ2002}. (1) is likely an oversimplification, we expect that some mixing of the phases may occur, particularly in the L-dominated fluid, however 
rather than a precise comparison, here we seek to identify a physical driving force for phase separation.

\section{Conclusion}

In summary, we report that suspensions of colloids of differing charge but near-equal size undergo phase separation into crystal and fluid phases, which is not expected on the basis of the usual Yukawa/DLVO description. Demixing appears to be initiated by crystallization of the strongly charged species and is stablised by one-body effects introduced by the small ions. The leading driving force is the ion-colloid interaction energy which favors partitioning to a higher density of strongly charged colloids.

Apart from our observation that crystallization of one species is involved, the mechanism of this novel phase separation in binary systems is open to investigation. An intriguing question concerns the fact that macroscopic phase separation appears to split into partially demixed regions of comparable volume: complete phase separation should result in a very asymmetric partitioning, to a low density phase of high-charge and high density phase of low-charge particles. It is possible that this is not reached due to kinetic trapping in the experiments.
Although we have offered what we believe to be the most compelling explanation, it would be extremely desirable to investigate these findings with more sophisticated theoretical treatments. In particular, at the two-body level, a full free-energy calculation which includes crystallisation of the strongly-charged species would be most desireable. In closing, we hope that the phenomenon of phase separation in binary like charged colloids will prove relevant to understanding of fluid-fluid phase separation in charged colloids and to segregation in biological systems and dusty plasmas.

\noindent \textbf{Acknowledgments} CPR gratefully acknowledges the Royal Society for funding. The authors are grateful to Jens Eggers, Bob Evans, Paul Hopkins, Rob Jack and Rene van Roij for helpful discussions.


\begin{thebibliography}{23}
\bibitem[{ren()}]{reneParams}
\bibinfo{note}{In the notation of ref. [YY], $A$ is set to 0 and
  $\kappa=\bar{\kappa}$.}
\end{thebibliography}

\end{document}